\documentclass[page-classic]{epl2}

\usepackage{amssymb}
\usepackage{amsmath}
\usepackage{amsopn}
\usepackage{graphics}
\usepackage{graphicx}
\usepackage{epsfig}
\usepackage{dsfont}

\shorttitle{M.-C. Firpo 2009 EPL \textbf{88} 30010}

\shortauthor{M.-C. Firpo 2009 EPL \textbf{88} 30010}

\institute{Laboratoire de Physique des Plasmas, CNRS-Ecole
Polytechnique, 91128 Palaiseau cedex, France}
{Phase transitions: general studies}     {Nonlinear
dynamics and chaos}     {Kinetic theory}
\abstract{Recently, there has been some vigorous interest in the
out-of-equilibrium quasistationary states (QSSs), with lifetimes
diverging with the number $N$ of degrees of freedom, emerging from
numerical simulations of the ferromagnetic XY Hamiltonian Mean Field
(HMF) starting from some special initial conditions. Phase
transitions have been reported between low energy magnetized QSSs
and large energy unexpected, antiferromagnetic-like, QSSs with low
magnetization. This issue is addressed here in the Vlasov
$N\rightarrow\infty$ limit. It is argued that the time asymptotic
states emerging in the Vlasov limit can be related to simple generic
time asymptotic forms for the force field. The proposed picture
unveils the nature of the out-of-equilibrium phase transitions
reported for the ferromagnetic HMF: This is a bifurcation point
connecting an effective integrable Vlasov one-particle
time-asymptotic dynamics to a partly ergodic one which means a
brutal open-up of the Vlasov one-particle phase space. Illustration
is given by investigating the time-asymptotic value of the
magnetization at the phase transition, under the assumption of a
sufficiently rapid time-asymptotic decay of the transient force
field. }

\begin{document}

\title{Unveiling the nature of out-of-equilibrium phase transitions in a
system with long-range interactions}

\shorttitle{EPL \textbf{88} 30010 (2009)} \shortauthor{EPL
\textbf{88} 30010 (2009)}
\author{M.-C. Firpo}
\date{M.-C. Firpo 2009 EPL \mathbf{88} 30010}
\maketitle

\section{Introduction}

Systems interacting via long-range interactions continue to receive a
considerable interest even in one dimension due to the intricate
relationships between their dynamical and statistical properties. Of
particular interest is the question of the convergence of the
time-asymptotic dynamics to equilibrium statistical mechanics ensemble
predictions. This issue is more than purely academic, it is relevant to
physical systems ranging from hot plasma physics and its ubiquitous
wave-particle interactions phenomena, including free electron lasers and
laser-plasma devices, to self-gravitating stellar systems. The Hamiltonian
Mean Field (HMF) model \cite{Antoni95}
\begin{equation}
H=\sum\limits_{i=1}^{N}\frac{p_{i}^{2}}{2}+\frac{c}{2N}\sum%
\limits_{i,j=1}^{N}\left[ 1-\cos \left( \theta _{i}-\theta _{j}\right) %
\right]  \label{HMF_Nfini}
\end{equation}%
is a choice toy model to explore those problems, since it exhibits a non
trivial long (and actually infinite) range collective dynamics meanwhile
permitting a simple exact derivation of the equilibrium statistical
mechanics. It describes the all-to-all interaction of $N$ particles moving
on the unit circle with momenta $p_{i}$ and canonically conjugated positions
$\theta _{i}$. Here $c=\pm 1$ contains the information on the attractive or
repulsive nature of the interaction: a positive $c$ gives the attractive
ferromagnetic model and a negative $c$ the repulsive antiferromagnetic one.

Recently, some unexpected out-of-equilibrium behaviors have been pointed
out: For the antiferromagnetic non-magnetized case, that is analogous to
plasma systems \cite{Elskens97}, and for low energies, the system
spontaneously develops into a biclustered state whose lifetime is an
increasing function of $N$ \cite{Barre2002,Leyvraz2002}. For the
ferromagnetic case, in which equilibrium statistical mechanics predicts a
second order phase transition, and for some special class of initial
conditions, the system has been shown to evolve to
out-of-thermal-equilibrium quasistationnary states (QSSs) with low
magnetization, still having lifetimes diverging with $N$, at subcritical
energies that would be thermodynamically associated to the magnetized phase.
This means that, depending on initial conditions, observables in the Vlasov $%
N\rightarrow \infty $ limit of the $N$-dimensional dynamics (\ref{HMF_Nfini}%
) do not converge asymptotically in time towards their equilibrium
statistical mechanics predictions: the limits $N\rightarrow \infty $ and $%
t\rightarrow \infty $ do not commute \cite{Tsallis99,Firpo2001,Pluchino07}.

The Vlasov equation that governs the evolution of the distribution function $%
f(p,\theta ,t)$ forms then the natural framework to investigate those QSSs.
For the HMF model in the ferromagnetic case ($c=1$), it reads
\begin{equation}
\frac{\partial f}{\partial t}+p\frac{\partial f}{\partial \theta }+E\left(
\theta ,t\right) \frac{\partial f}{\partial p}=0,  \label{Vlasov_HMF}
\end{equation}%
where $E$ stands for the force field given by
\begin{equation}
E\left( \theta ,t\right) =-\int\limits_{-\infty }^{+\infty
}du\int\limits_{0}^{2\pi }d\alpha \sin (\theta -\alpha )f\left( u,\alpha
,t\right) .  \label{Vlasov_E_eq1}
\end{equation}%
Eq. (\ref{Vlasov_E_eq1}) forms a consistency relation that makes Vlasov
equation (\ref{Vlasov_HMF}) intrinsically nonlinear. It is meaningful to
introduce the magnetization vector $\mathbf{M}=(M_{x},M_{y})$ through $M_{x}%
\left[ f\right] =\int \int dud\alpha \cos \alpha f\left( u,\alpha ,t\right) $
and $M_{y}\left[ f\right] =\int \int dud\alpha \sin \alpha f\left( u,\alpha
,t\right) $, which enables to write the one-particle Hamiltonian associated
to (\ref{Vlasov_HMF}) as
\begin{equation}
h_{1}\left( p,\theta ,t\right) =\frac{p^{2}}{2}-M_{x}\left[ f\right] (t)\cos
\theta -M_{y}\left[ f\right] (t)\sin \theta .  \label{def_h1}
\end{equation}

The first aim of this Letter is to propose that the time asymptotic
states emerging in the Vlasov limit be related to simple generic
time asymptotic forms for the force field $E\left( \theta ,t\right)
$. The proposed picture unveils the nature of the out-of-equilibrium
phase transitions reported for the ferromagnetic HMF: This is a
bifurcation point connecting an integrable Vlasov time-asymptotic
one-particle dynamics to a partly non-integrable and ergodic one.
Illustration is given by investigating the time-asymptotic value of
the magnetization at the phase transition, under the assumption of a
sufficiently rapid time-asymptotic decay of the transient force
field. Then, a proof of principle is given, that the time-asymptotic
state of the system can be completely derived as an initial value
problem, in the domain where initial magnetizations are sufficiently
large that the out-of-equilibrium phase transitions be of second
order. This serves to specify the validity regime of the proposed
time-asymptotic picture.

\section{Numerical evidence}

Considering the ferromagnetic Hamiltonian (\ref{HMF_Nfini}) with $c=1$, it
can be shown \cite{Antoni95} that the system undergoes a second order phase
transition for $U\equiv H/N=\varepsilon _{c}=3/4$ with order parameter the
modulus of the magnetization $M\equiv \left\Vert \mathbf{M}\right\Vert $.
QSSs have been observed starting with initial waterbag distributions of
particle momenta and positions of the form%
\begin{equation}
f_{0}(p,\theta )=\frac{1}{4\Delta p\Delta \theta }\mathds{1}_{[-\Delta
p;\Delta p]}(p)\mathds{1}_{[-\Delta \theta ;\Delta \theta ]}(\theta ),
\label{initial_f0_WB}
\end{equation}%
where $\mathds{1}_X$ denotes the characteristic function of the domain $X$, $%
\Delta p\geq 0$ and $0\leq \Delta \theta \leq \pi $. There is then a
one-to-one relationship between $\left( \Delta p,\Delta \theta \right) $ and
$\left( U,M_{0}\right) $, where $M_{0}$ is the initial magnetization,
through $M_{0}=\sin \left( \Delta \theta \right) /\Delta \theta $ and $%
U=\Delta p^{2}/6+(1-M_{0}^{2})/2$. To be specific, a transition between two
sorts of time-asymptotic states has been reported. For a given $\Delta
\theta $, that is for a given initial magnetization $M_{0}$, there exists
some critical energy density $U_{c}$, whose value is a growing function of $%
M_{0}$, such that: For $U<U_{c}(M_{0})$, the system converges asymptotically
towards the clustered state as predicted by equilibrium statistical
mechanics whereas, for $U>U_{c}(M_{0})$, magnetization drops and the system
mimics an antiferromagnetic (upper critical) behavior in the limit of large $%
N$. This latter regime displays a specific signature: whereas, for $%
U<U_{c}(M_{0})$, the one-particle phase space is that of the usual single
resonance pendulum centered on $p=0$ associated to the clustered phase, for $%
U>U_{c}(M_{0})$, it shows a superposition of two waves traveling at opposite
phase velocities. This has been enlightened very recently through very
detailed numerical simulations in Ref. \cite{Bachelard08} (see also Ref.
\cite{AntoniazziPRL2007} for large time one-particle phase space plots
obtained from Vlasov simulations of the HMF model in the low-magnetized
phase).

These out-of-equilibrium phase transitions have been classified in
two types: for a sufficiently low initial magnetization, below some
threshold value $M_{0t} \approx 0.15$, transitions are of the first
order type with the order parameter, namely the magnetization, being
discontinuous at the critical point. For a larger initial
magnetization $M_{0}>M_{0t}$, transitions are of second order with
no discontinuity of the magnetization at the critical point. This
has been reported in extensive finite-$N$ numerical simulations in
Ref. \cite{Antoniazzi2007} and the agreement with statistical
predictions by P.H. Chavanis \cite{Chavanis2006} following
Lynden-Bell's approach has proved to be remarkably good. However, a
careful observation of QSS magnetizations in Fig. 2 shows that, at
least in the second order regime of out-of-equilibrium phase
transitions for $M_{0}>M_{0t}$ (bottom plot), its upper critical
values appear to remain strictly positive even in the $N \rightarrow
\infty$ limit. This means that there is a continuity of the order
parameter at the transition point yet this does not vanish in the
high energy phase. This is absolutely consistent with the above
mentioned QSS phase space pictures. Actually, the observed long
time, upper critical, phase space inhomogeneity reflected by the two
contra-propagating clusters implies that the modulus of the
magnetization is, even small, strictly positive and that particles
do not behave asymptotically in time as free particles even in the
Vlasov limit.

This latter behavior belongs to a more general phenomenology. It is similar
to the one emerging commonly from long time simulations of
antiferromagnetic-like 1D Vlasov-Poisson systems \cite{Demeio}. Unless the
electric field is totally damped to zero, there is indeed a shared agreement
that time-asymptotic states can be represented as a superposition of
traveling waves \cite{LancelloPRL98,LancellottiDorning03}. The physical
picture behind this is that nonlinearities, that eventually come into play,
take place in the form of particle trapping which freezes the dynamics.

\section{Time-asymptotic bifurcation}

By analogy with the almost periodic time-asymptotic states observed
in plasma Vlasov-Poisson systems, it is possible to infer that the
time asymptotic form of the force field, $A$, may be put in the form
of a sum of traveling waves. Having this in mind, the time
asymptotic force field corresponding to the usual magnetized
clustered phase, with $U<U_{c}(M_{0})$, may be written as a
time-independent (zero phase velocity) single wave
\begin{equation}
A(\theta )=-a\sin \left( \theta -\varphi \right) .  \label{Aferro}
\end{equation}%
On the contrary, for $U>U_{c}(M_{0})$, the system displays a typical
plasma-like (i.e. antiferromagnetic) behavior and the time
asymptotic force field may be written as a superposition of two
traveling
waves with opposite phase velocities $\omega > 0$ and $-\omega $%
\begin{equation}
A\left( \theta ,t\right) =-\frac{a}{2}\sin \left( \theta -\varphi +\tilde{%
\varphi}-\omega t\right) -\frac{a}{2}\sin \left( \theta -\varphi -\tilde{%
\varphi}+\omega t\right).  \label{Aantiferro}
\end{equation}%
It can be simply checked \cite{FirpoUnpub} that this two-wave form (\ref%
{Aantiferro}) is the minimal superposition of traveling waves compatible
with the HMF cosine potential (\ref{Vlasov_E_eq1}). Here $\varphi $ and $%
\tilde{\varphi}$ represent constant angles and $a>0$ is the constant wave
amplitude. All these parameters ($a$, $\omega $, $\varphi $ and $\tilde{%
\varphi}$) relate to the initial conditions through $U$ and $M_{0}$, yet
this dependence has not been made explicit to ease notations. In order to
continuously match (\ref{Aferro}) and (\ref{Aantiferro}) at the transition
point, we should take $\tilde{\varphi}=0$ and $\omega =0^{+}$ in (\ref%
{Aantiferro}) when $U=U_{c}(M_{0})$. Eqs. (\ref{Aferro}) and (\ref%
{Aantiferro}) capture, so to speak, the asymptotic \textit{skeleton} of the
Vlasov HMF dynamics evolving from the initial conditions (\ref{initial_f0_WB}%
).

The continuity of $\omega$ at the transition is an assumption
compatible with available numerical simulations undertaken in the
second order regime of the out-of-equilibrium phase transitions. For
instance, the bifurcation diagram plotted in Fig. 3 of Ref.
\cite{Bachelard08} shows that the phase transition is signaled by a
pinch in the velocity space. As the energy increases as one goes
right from this pinch, two resonances first fully overlap then
gradually emerge in a symmetric way. This is further commented in
the concluding discussion. A
scheme of this time-asymptotic bifurcation is depicted in Fig. \ref%
{Fig_skeleton}. The out-of-equilibrium phase transition is then associated
to the adiabatic limit $\omega \rightarrow 0$ in Eq. (\ref{Aantiferro}) that
separates an integrable one-particle dynamics for the condensed phase to a
one-and-a-half degrees of freedom (d.o.f.) one-particle dynamics for the
antiferromagnetic-like phase \cite{note}.

\begin{figure}[tbp]
\includegraphics[width=6 cm, angle=-90]{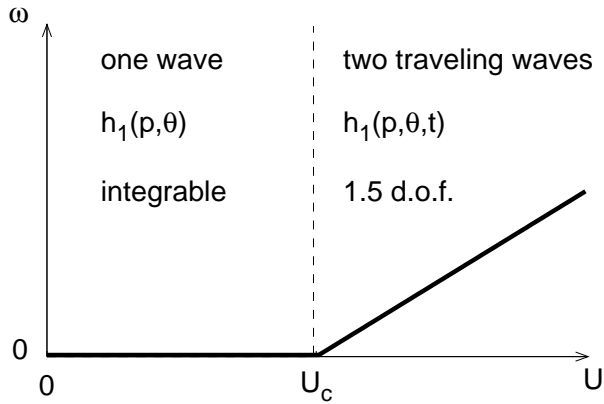}
\caption{Schematic view of the out-of-equilibrium bifurcation in terms of
wave pulsation between asymptotic ferro- and antiferromagnetic states. Only
the $\protect\omega \geq 0$ half-space has been represented.}
\label{Fig_skeleton}
\end{figure}

\section{Analysis}

Investigating the time asymptotic fate of the system as an initial value
problem forms a formidable task, because of the nonlinearity of the Vlasov
equation induced by (\ref{Vlasov_E_eq1}), that is out of the scope of the
present paper. Instead, we wish to investigate further the implications of
the picture presented in Fig. \ref{Fig_skeleton} to unveil the dynamical
nature of the out-of-equilibrium phase transition. To do so, let us follow
briefly the approach proposed by Lancellotti and Dorning \cite%
{LancelloPRL98,LancellottiDorning03} for the 1D Vlasov-Maxwell system, and
decompose the force field $E$ into its time-asymptotic part $A$ and a
transient part $T$ such that
\begin{equation}
E\left( \theta ,t\right) =A\left( \theta ,t\right) +T\left( \theta ,t\right)
,  \label{decomposition_E}
\end{equation}%
with $\lim_{t\rightarrow \infty }T\left( \theta ,t\right) =0$. Our
interest lies, in particular, in the time asymptotic value of the
magnetization, that is a collective observable, in the vicinity of
the phase transition. Due to the decomposition
(\ref{decomposition_E}) it is possible to extract $a$ in Eqs.
(\ref{Aferro}) and (\ref{Aantiferro}) from suitable time averages.
This can be straightforwardly done through
\begin{equation}
-\frac{a}{4}\left( 1+\delta _{\omega ,0}\right) =\lim_{t\rightarrow \infty }%
\frac{1}{2\pi t}\int\limits_{0}^{t}d\tau \int\limits_{0}^{2\pi }d\theta
E\left( \theta ,\tau \right) \sin \left( \theta -\varphi \right) \cos \left(
\tilde{\varphi}-\omega \tau \right) ,  \label{def_a1}
\end{equation}%
that gives, using (\ref{Vlasov_E_eq1}) and (\ref{decomposition_E});,%
\begin{equation}
\frac{a}{2}\left( 1+\delta _{\omega ,0}\right) =\left\langle
\int\limits_{-\infty }^{+\infty }du\int\limits_{0}^{2\pi }d\alpha f\left(
u,\alpha ,\tau \right) \cos \left( \tilde{\varphi}-\omega \tau \right) \cos
\left( \alpha -\varphi \right) \right\rangle _{\tau },  \label{def_a}
\end{equation}%
where $\left\langle .\right\rangle _{\tau }\equiv \lim_{t\rightarrow \infty }%
\frac{1}{t}\int\nolimits_{0}^{t}d\tau $ denotes the time average. In
Eqs. (\ref{def_a1}) and (\ref{def_a}), we used
$\tilde{\varphi}(\omega =0)=0$. As previously mentioned, this,
together with the continuity of $\omega$ at the transition, can be
called a "second order type" assumption in the sense that one moves
continuously from the low energy to the high energy phases at the
critical point.

In order to relate the time-asymptotic characterization to the
initial state, a rather crude assumption will be done since the
contribution coming from the transient force-field will be
discarded. Practically, this amounts to consider that the decay of
the transient force-field is so fast that the form of the particle
distribution function does not deviate substantially from that of
the initial condition, so that one can write for any $\left(
p,\theta ,t\right)$
\begin{equation}
f\left( p,\theta ,t\right) \simeq f_{0}\left[ p_{0}^{A}\left( p,\theta
,t\right) ,\theta _{0}^{A}\left( p,\theta ,t\right) \right],
\label{formal_solution}
\end{equation}
where $f_{0}\left( p,\theta \right) =f\left( p,\theta ,t=0\right) $ stands
for the initial distribution function and where $\left[ p_{s}^{A}\left(
p,\theta ,t\right) ,\theta _{s}^{A}\left( p,\theta ,t\right) \right] $
denotes the phase space location at time $s$ of a particle arriving in $%
(p,\theta )$ at time $t$ under the time-asymptotic one-particle
dynamics. This derives from the one-particle time-asymptotic
Hamiltonian $h_{1}(p,\theta ,t)$ through the characteristics
$d\theta /dt=p\equiv
\partial h_{1}/\partial p$ and $dp/dt=A\left( \theta ,t\right) \equiv
-\partial h_{1}/\partial \theta $.

\section{Nature of the out-of-equilibrium phase transition}

From now on, we focus on the $\omega=0$ limit that is associated to the
out-of-equilibrium phase transition and look for an illustration of the
implication of its dynamical signature as depicted on Fig. \ref{Fig_skeleton}%
. Due to Eq. (\ref{formal_solution}), the identity (\ref{def_a}) becomes
simply
\begin{equation}
a=\left\langle \int\limits_{-\infty }^{+\infty }du\int\limits_{0}^{2\pi
}d\alpha f_{0}\left[ u_{0}^{A}\left( u,\alpha ,\tau \right) ,\alpha
_{0}^{A}\left( u,\alpha ,\tau \right) \right] \cos \left( \alpha -\varphi
\right) \right\rangle _{\tau }.  \label{def_a_omega0}
\end{equation}%
Using Liouville's theorem on the conservation of phase space and Eq. (\ref%
{initial_f0_WB}) and inverting the order of the time integration, this reads
\begin{equation}
a=\frac{1}{4\Delta p\Delta \theta }\int\limits_{-\Delta p}^{+\Delta
p}du_{0}\int\limits_{-\Delta \theta }^{\Delta \theta }d\alpha
_{0}\left\langle \cos \left( \alpha -\varphi \right) \right\rangle _{\tau },
\label{eval_a_omega0}
\end{equation}%
where the index $A$ has been dropped to ease notations.

\subsection{Limit $\protect\omega \rightarrow 0^{-}$}

Let us first evaluate (\ref{eval_a_omega0}) from the left (integrable) side.
Then, since $A$ does not depend on time, the one-particle Hamiltonian $%
h_{1}(p,\theta )=p^{2}/2-a\cos \left( \theta -\varphi \right) $ is
integrable, and the time average in (\ref{def_a_omega0}) can be replaced
here by an ensemble average on the energy level $h_{1}(u,\alpha
)=h_{1}(u_{0},\alpha _{0})=E_{0}$. Let us denote by $\Theta (x)$ the usual
Heaviside function defined by $\Theta (x)\equiv \int\nolimits_{-\infty
}^{x}dy\delta (y)$ and by $Z(a,E)$ the one-particle volume partition
function defined by
\begin{equation}
Z(a,E)\equiv \int\limits_{-\infty }^{\infty }du\int\limits_{-\pi }^{\pi
}d\alpha \Theta \left[ E-h_{1}(u,\alpha )\right] .
\label{def_partition_volume}
\end{equation}%
Then, it is easy to check that
\begin{equation}
\left\langle \cos \left( \alpha -\varphi \right) \right\rangle _{\tau }=%
\frac{1}{2}\left. \frac{\partial _{a}Z}{\partial _{E}Z}\right\vert
_{(a,E_{0})}.  \label{def_ensemble_average_cos}
\end{equation}%
For the trapped motion $-a\leq E_{0}\leq a$, moving to action-angle
variables yields $Z(a,E)=\int \int dJd\psi \Theta \left[ E-h_{1}(J)\right]
=2\pi \int dh\omega (h)^{-1}\Theta \left( E-h\right) $ with $\omega (h)=\pi
\sqrt{a}/(2K(k))$ \cite{LichtenbergLiberman} where $k=$ $\sqrt{\left(
h+a\right) /2a}$ and $K(k)=\int\nolimits_{0}^{\pi /2}d\phi /\sqrt{1-k\sin
^{2}\phi }$, the complete elliptic integral of the first kind. This gives,
using (\ref{def_ensemble_average_cos}),%
\begin{equation}
\left\langle \cos \left( \alpha -\varphi \right) \right\rangle _{\tau }=%
\frac{2E\left( \frac{E_{0}+a}{2a}\right) }{K\left( \frac{E_{0}+a}{2a}\right)
}-1,  \label{moy_cos_piegeage}
\end{equation}%
where $E(k)=\int\nolimits_{0}^{\pi /2}d\phi \sqrt{1-k\sin ^{2}\phi }$ is the
complete elliptic integral of the second kind. For the untrapped motion $%
E_{0}>a$, one has directly $Z(a,E)=2\sqrt{2}\int\nolimits_{-\pi }^{\pi
}d\alpha \sqrt{E_{0}+a\cos \left( \alpha -\varphi \right) }$. This yields%
\begin{equation}
\left\langle \cos \left( \alpha -\varphi \right) \right\rangle _{\tau }=%
\frac{\left( \frac{E_{0}}{a}+1\right) E\left( \frac{2a}{a+E_{0}}\right) }{%
K\left( \frac{2a}{a+E_{0}}\right) }-\frac{E_{0}}{a}.  \label{moy_cos_libre}
\end{equation}%
It remains to evaluate the double integral (\ref{eval_a_omega0}) using the
expressions (\ref{moy_cos_piegeage}) for $-a\leq E_{0}\leq a$ and (\ref%
{moy_cos_libre}) for $E_{0}>a$, with $E_{0}=u_{0}^{2}/2-a\cos \left(
\alpha_{0} -\varphi \right)$.

\subsection{Limit $\protect\omega \rightarrow 0^{+}$}

Let us now consider the limit $\omega \rightarrow 0^{+}$ of the 1.5 degrees
of freedom one-particle Hamiltonian with force field given by Eq. (\ref%
{Aantiferro}). This can be written as $A(\theta ,t)=-a\cos \left( \tilde{%
\varphi}-\omega t\right) \sin \left( \theta -\varphi \right) $. This
corresponds to the pendulum with variable amplitude. The essential thing to
recognize here is that the limit $\omega \rightarrow 0$ is \textit{singular}%
: Actually, because of separatrix crossings, the 1.5 d.o.f. Hamiltonian
associated to (\ref{Aantiferro}) exhibits an ergodic, diffusive dynamics on
a confined phase space region even in the adiabatic limit $\omega
\rightarrow 0^{+}$. The ergodic region corresponds to the phase space region
swept by the resonance cat's eye in which particles transit between trapped
and untrapped motion \cite{Menyuk85,ElskensEscande,MirbachCasati99}.
Consequently, in the limit $\omega \rightarrow 0^{+}$, the evaluation of $a$
in Eq. (\ref{eval_a_omega0}) differs from the previous $\omega \rightarrow
0^{-}$ calculation because of a different contribution from the initial
phase space domain swept by the separatrix, i.e. such that $-a\leq
h_{1}(u_{0},\alpha _{0})\leq a$. In this ergodic domain, $\left\langle \cos
\left( \alpha -\varphi \right) \right\rangle _{\tau }$ is given by its
average on the inner whole cat's eye region
\begin{equation}
\left\langle \cos \left( \alpha -\varphi \right) \right\rangle _{\tau }=%
\frac{\int\limits_{-\pi }^{\pi }d\alpha \cos \left( \alpha -\varphi \right)
\sqrt{1+\cos \left( \alpha -\varphi \right) }}{\int\limits_{-\pi }^{\pi
}d\alpha \sqrt{1+\cos \left( \alpha -\varphi \right) }}=\frac{1}{3}.
\label{moy_cos_ergodic}
\end{equation}
The outer region, for which $h_{1}(u_{0},\alpha _{0})> a$, corresponds to a
regular one-particle motion and gives the same contribution to (\ref%
{eval_a_omega0}) than in the $\omega \rightarrow 0^{-}$ case. In evaluating
the double integral (\ref{eval_a_omega0}) in the $\omega \rightarrow 0^{+}$
limit, one should then use the identity (\ref{moy_cos_libre}) for $E_{0}>a$,
while using the expression (\ref{moy_cos_ergodic}) for $-a\leq E_{0}\leq a$.

\subsection{Implications}

The $\omega \rightarrow 0^{-}$ and $\omega \rightarrow 0^{+}$ values of the
time-asymptotic magnetization $a$ were numerically computed.
\begin{figure}[htbp]
\includegraphics[width=8.8 cm]{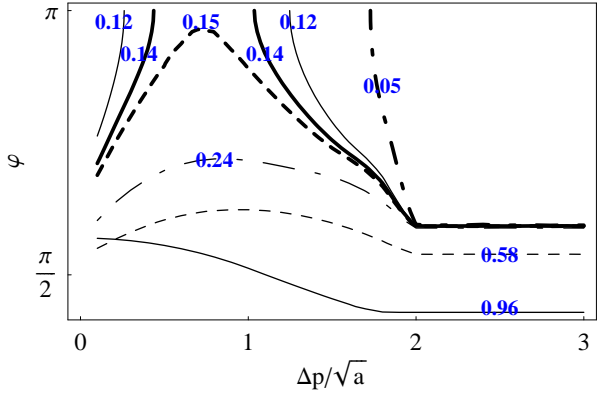}
\caption{Curves in the $(\Delta p/\protect\sqrt{a},\protect\varphi)$ space
on which $a(\protect\omega \rightarrow 0^{-})=a(\protect\omega \rightarrow
0^{+})$ labeled by the values of the initial magnetization $M_{0}$. The
range of the angle $\protect\varphi$ is restricted to the interval $[0;%
\protect\pi]$.}
\label{Fig_OequalOplus}
\end{figure}
Figure \ref{Fig_OequalOplus} represents the curves, in the $(\Delta p/\sqrt{a%
},\varphi)$ space, on which the time asymptotic magnetizations $a$ for $%
\omega = 0^{-}$ and $\omega = 0^{+}$ coincide, for several values of the
initial magnetization $M_{0}$. Here the range of the angle $\varphi$ is
restricted to the interval $[0;\pi]$. The full plot in the $[0;2\pi]$ range
can be recovered by an axial symmetry about the horizontal line $\varphi=\pi$
(since when $\varphi$ is solution, then $2 \pi - \varphi$ is also solution
by switching the up and down waves in (\ref{Aantiferro})). When $\Delta p/%
\sqrt{a}$ becomes larger than 2, $\varphi$ remains constant since resonance
cat's eye has been fully covered and that regular contributions (\ref%
{moy_cos_libre}) from the outer region cancel out.

Interestingly enough, when $M_{0}$ is sufficiently small (roughly below
0.15), there appears a gap, namely a region of the rescaled parameter $%
\Delta p/\sqrt{a}$, where the time asymptotic magnetizations for $\omega
\equiv 0$ and $\omega = 0^{+}$ do not intersect. This is reminiscent of the
first order nature of the out-of-equilibrium phase transitions, associated
with a discontinuity of the order parameter at the transition, evidenced in
some already mentioned previous numerical works \cite%
{Antoniazzi2007} and thermodynamical analysis \cite{Chavanis2006}, for the
same domain of $M_{0}$. It is interesting to note here that the analysis by
itself is able to unveil that the presumed continuity between Eqs. (\ref%
{Aferro})-(\ref{Aantiferro}) cannot be satisfied in that domain, since $a$
is not continuous.

\section{Derivation of the time-asymptotic state: a proof
of principle}

In order to fully characterize the transition, namely, for a given
$M_{0}$, determine the time-asymptotic values of the magnetization
$a$, of the wave phase $\varphi $ and determine the transition
energy through $\Delta p$, an extra condition is needed. The natural
additional relation to fulfill is
given by the continuity of the energy at the transition. Its expression is $%
h[f]\equiv K[f]+1/2\left[ 1-\left( M[f]\right) ^{2}\right] $, with the
kinetic energy $K[f]=\int \int p^{2}/2f\left( p,\theta ,t\right) dpd\theta $%
. In the regular domain, namely for the case $\omega \rightarrow 0^{-}$ with
$h_{1}(p,\theta )=h_{1}\left( p_{0},\theta _{0}\right) =p^{2}/2-a\cos \left(
\theta -\varphi \right) $, neglecting the effect of transient and using
Liouville theorem, the time average of the kinetic energy is given by%
\begin{eqnarray}
\left\langle K[f]\right\rangle _{t}^{\omega \rightarrow 0^{-}} &=&\frac{1}{%
4\Delta p\Delta \theta }\int\limits_{-\Delta p}^{+\Delta
p}dp_{0}\int\limits_{-\Delta \theta }^{\Delta \theta }d\theta
_{0}h_{1}\left( p_{0},\theta _{0}\right) +\frac{a}{4\Delta p\Delta \theta }%
\int\limits_{-\Delta p}^{+\Delta p}dp_{0}\int\limits_{-\Delta \theta
}^{\Delta \theta }d\theta _{0}\left\langle \cos \left( \theta -\varphi
\right) \right\rangle _{t}  \notag \\
&=&\frac{1}{2\Delta p}\int\limits_{-\Delta p}^{+\Delta p}dp_{0}\frac{%
p_{0}^{2}}{2}-\frac{a}{2\Delta \theta }\int\limits_{-\Delta \theta }^{\Delta
\theta }d\theta _{0}\cos \left( \theta _{0}-\varphi \right) +a^{2},
\label{moy_energie_cinetique}
\end{eqnarray}%
where the identity (\ref{eval_a_omega0}) was used. The continuity of the
energy at the phase transition amounts then to the identity%
\begin{equation}
\frac{\Delta p^{2}}{6}-aM_{0}\cos \varphi +a^{2}-\left\langle
K[f]\right\rangle _{t}^{\omega \rightarrow 0^{+}}=\frac{1}{2}\left(
\left\langle M[f]^{2}\right\rangle _{t}^{\omega \rightarrow
0^{-}}-\left\langle M[f]^{2}\right\rangle _{t}^{\omega \rightarrow
0^{+}}\right) ,  \label{add_relation}
\end{equation}%
where time averages may be computed using the exact characteristics for $%
h_{1}$ in the $\omega \rightarrow 0^{-}$ case and the numerically
obtained ones in the $\omega \rightarrow 0^{+}$ case.

\begin{figure}[tbph]
\includegraphics[width=8.8 cm]{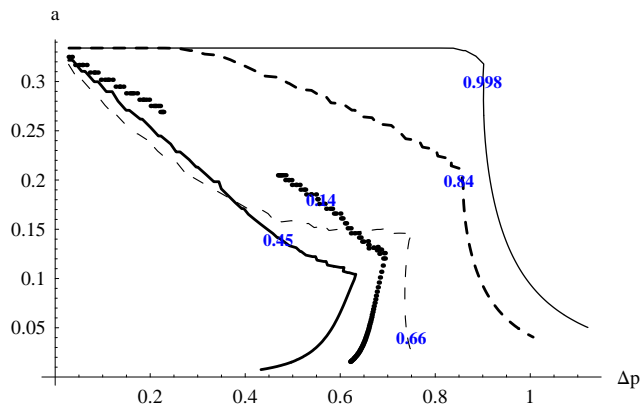}
\caption{Curves in the $(\Delta p,a)$ space on which $a(\protect\omega %
\rightarrow 0^{-})=a(\protect\omega \rightarrow 0^{+})$ labeled by the
values of the initial magnetization $M_{0}$.}
\label{Fig_ATrans}
\end{figure}

This procedure would amount to determine the transition as an
initial value problem, a truly desirable perspective that would
involve the knowledge of the sole one-particle Vlasov dynamics
without any reference to equilibrium statistical mechanics. Yet this
desirable perspective is out of reach of the present analysis,
surely due to the presently too stringent condition on the decay of
the transient force field. The obtained values of the transition
energies \cite{FirpoUnpub} are actually sensibly below the
numerically \cite{Bachelard08} or thermodynamically computed ones
\cite{Chavanis2006}. This discrepancy can be already inferred from
Fig. \ref{Fig_ATrans}. This Figure displays the possible
time-asymptotic values of the magnetization $a$ at the phase
transition in the case where $a(\omega \rightarrow 0^{-})$ and
$a(\omega \rightarrow 0^{+})$ coincide. As in Fig.
\ref{Fig_OequalOplus}, it is clear that, for low enough $M_{0}$,
there exists some forbidden range for $\Delta p$ that may signal a
discontinuity of the order parameter associated to the first order
transition region. It is also clear from this picture, that, at
least for low enough $M_{0}$, the possible $\Delta p$ are bounded by
values ($\simeq 1$) which would yield values of the energy
transition $U_{c}(M_{0})$ below the available numerically computed
\cite{Antoniazzi2007} and thermodynamically predicted ones
\cite{Chavanis2006}.
Finally, in the domain of initial magnetizations where the
out-of-equilibrium phase transitions are of first order, the
additional identity (\ref{add_relation}) is insufficient to close
the system of time-asymptotic parameters because $a$ is not
continuous at the transition. This signals that the continuity
assumption made in Eqs. (\ref{Aferro}) and (\ref{Aantiferro})
breaks:  $\omega(U,M_{0})$ is strictly positive in the upper
critical domain and its knowledge at the "right side" of transition
line is required.

\section{Conclusion}

To the author's knowledge, this study is the first to address the
issue of out-of-equilibrium phase transitions in the HMF model from
the perspective of an initial value problem in the Vlasov limit. The
essential point of this Letter is to propose a picture unveiling the
nature of the phase transition: This corresponds to a brutal open-up
of the time-asymptotic Vlasov one-particle phase space. The phase
transition coincides with a jump between a time-asymptotic phase
space foliated by energy lines and a phase space divided between an
ergodic and a regular \cite{note2} components. It is proposed as a
mechanism for second order phase transitions compatible with
\textit{non-vanishing} time-asymptotic values of the order parameter
in mean-field long-range systems. As a byproduct, this study shows
indirectly that the a priori reasonable hypothesis of a rapid decay
of the time-asymptotic transient force field may not be satisfied by
the HMF model. This is a posteriori not astonishing in view of the
numerous recent papers reporting incomplete relaxation and
deficient mixing properties in such a long-range system \cite%
{Tamarit2005,Jain2007,YYY2007,Campa2008,Figueiredo08}. It remains to
elucidate whether relaxing the hypothesis on the transient field
would still be compatible with a purely dynamical initial value
treatment, e.g. in the spirit of Lancellotti and Dorning's approach.
 The present approach underlines the importance of the knowledge of
the time-asymptotic spectrum $\omega(U,M_{0})$ of the dynamics. It
would be instructive to proceed to time-Fourier transforms of large
time Vlasov numerical simulations in order to extract this
information.

Very recently, the author became aware of the just published paper by
Leoncini \textit{et al.} \cite{Leoncini2009} whose possible connections with
the present approach would be interesting to explore.

\acknowledgments

MCF thanks A.F. Lifschitz for computational support. A fruitful
discussion with D. Fanelli is gratefully acknowledged.


\begin{thebibliography}{99}
\bibitem{Antoni95} \Name{Antoni M. \and Ruffo S.} \REVIEW{Phys. Rev. E} {52}{%
1995}{2361}.

\bibitem{Elskens97} \Name{Elskens Y. \and Antoni M.}
\REVIEW{Phys. Rev.
E}{55}{1997}{6575}.

\bibitem{Barre2002} \Name{Barr\'{e} J., Bouchet F., Dauxois T., Ruffo S.}
\REVIEW{Eur.
Phys. J. B}{29}{2002}{577}.

\bibitem{Leyvraz2002} \Name{Leyvraz F., Firpo M.-C., Ruffo S.}
\REVIEW{J. Phys.
A}{35}{2002}{4413}.

\bibitem{Tsallis99} \Name{Tsallis C.} \REVIEW{Brazil. J. Phys.}{29}{1999}{1}.

\bibitem{Firpo2001}
\Name{Firpo M.-C., Doveil F., Elskens Y., Bertrand P., Poleni M. \and Guyomarc'h
D.} \REVIEW{Phys. Rev. E}{64}{2001}{026407}.

\bibitem{Pluchino07} \Name{Pluchino A., Rapisarda A. \and Tsallis C.} %
\REVIEW{EPL}{80}{2007}{26002}.

\bibitem{Bachelard08}
\Name{Bachelard R., Chandre C., Fanelli D., Leoncini X., \and
Ruffo S.} \REVIEW{Phys. Rev. Lett.}{101}{2008}{260603}.

\bibitem{AntoniazziPRL2007} \Name{Antoniazzi A., Califano F., Fanelli D.
\and Ruffo S.} \REVIEW{Phys. Rev. Lett.}{98}{2007}{150602}.

\bibitem{Antoniazzi2007}
\Name{Antoniazzi A., Fanelli D., Ruffo S.
\and Yamaguchi Y.Y.} \REVIEW{Phys. Rev. Lett.}{99}{2007}{040601}.

\bibitem{Chavanis2006} \Name{Chavanis P.H.}
\REVIEW{Eur.
Phys. J. B}{53}{2006}{487}.

\bibitem{FirpoUnpub} \Name{Firpo M.-C.} unpublished.

\bibitem{Demeio} \Name{Demeio L. \and Holloway J.P.}
\REVIEW{J.
Plasma Physics}{46}{1991}{63}.

\bibitem{note} It cannot, also, be definitely ruled out, on the basis of
recent finite-$N$ numerical simulations \cite{Bachelard08}, that the
proposed time-asymptotic picture, as depicted in Fig. \ref{Fig_skeleton}, be
embedded in additional periodicities, such as a time-asymptotic periodic
behavior of the magnetization. Vlasov simulations, in the line of Ref. \cite%
{AntoniazziPRL2007}, would be required to ascertain this by discarding
finite-$N$ effects. This would not, however, change the phase transition
description that considers only the time average $\omega=0$ limit.

\bibitem{LancelloPRL98} \Name{Lancellotti C. \and Dorning J.J.}
\REVIEW{Phys. Rev.
Lett.} {81}{1998}{5137}.

\bibitem{LancellottiDorning03} \Name{Lancellotti C. \and Dorning J.J.}
\REVIEW{Phys. Rev.
E} {68}{2003} {026406}.

\bibitem{LichtenbergLiberman} \Name{Lichtenberg A.J. \and Lieberman M.A.} %
\Book{Regular and Stochastic Motion} \Publ{Springer-Verlag, New York} %
\Year{1983}.

\bibitem{Menyuk85} \Name{Menyuk C.R.} \REVIEW{Phys. Rev. A} {31}{1985}{3282}.

\bibitem{ElskensEscande} \Name{Elskens Y. \and Escande D.F.} \REVIEW{Physica
D}{62}{1993}{66}.

\bibitem{MirbachCasati99} \Name{Mirbach B. \and Casati G.} \REVIEW{Phys.
Rev. Lett.}{83}{1999}{1327}.

\bibitem{note2} Integrability here is relative to the time-asymptotic limit. This
does not prevent the dynamics in the low energy domain to exhibit
some ergodic properties through the transient field and does not
invalidate statistical mechanics.

\bibitem{Tamarit2005}
\Name{Tamarit F.A., Maglione G., Stariolo D.A.
\and Anteneodo C.} \REVIEW{Phys. Rev. E}{71}{2005}{036148}.

\bibitem{Jain2007} \Name{Jain K., Bouchet F. \and Mukamel D.} \REVIEW{J.
Stat. Mech.}{}{2007}{P11008}.

\bibitem{YYY2007} \Name{Yamaguchi Y.Y., Bouchet F. \and Dauxois T.} %
\REVIEW{J. Stat. Mech.}{}{2007}{P01020}.

\bibitem{Campa2008}
\Name{Campa A., Chavanis P.H., Giansanti A., \and
Morelli G.} \REVIEW{Phys. Rev. E}{78}{2008}{040102(R)}.

\bibitem{Figueiredo08} \Name{Figueiredo A., Rocha Filho T.M. \and Amato M.A.}
\REVIEW{EPL}{83}{2008}{30011}.

\bibitem{Leoncini2009} \Name{Leoncini X., Van Den Berg T.L. and Fanelli D.} %
\REVIEW{EPL}{86}{2009}{20002}.


\end{thebibliography}
\end{document}